# Eco-friendly Bismuth Based Double Perovskites $X_2NaBiCl_6$ (X=Cs, Rb, K) for Optoelectronic and Thermoelectric Applications: A First-Principles Study.

Syed Zuhair Abbas Shah[*,1,2], Shanawer Niaz[2,3], Tabassum Nasir[1], and James Sifuna[4,5]

[1]Institute of Physics, Gomal University, D.I. Khan, 29050, Pakistan

[2]Department of Physics, Thal University Bhakkar, Bhakkar, 30000, Pakistan

[3]Molecular Engineering Laboratory, at the Department of Physics, University of Patras, Patras GR-26500, Greece

[4]Department of Natural Science, The Catholic University of Eastern Africa, 62157 - 00200, Nairobi, Kenya.

[5]Department of Technical and Applied Physics, The Technical University of Kenya, 52428-00200, Nairobi, Kenya.

*Email: zuhair.abbas@uos.edu.pk

*Tel: +92-300-7668414

## Abstract

Owing to the energy shortages and various severe adverse effects of traditional fossil fuel power generation mechanisms, photovoltaic and thermoelectric materials are considered as the potential candidates for building non-traditional, efficient, and eco-friendly power generation portfolios. Lead-based perovskites have emerged as highly efficient, abundantly available, and low-cost materials for such applications but there are two major challenges i.e. chemical instability and the danger of toxic lead leaching that can badly affect both the environment and human health. Therefore, in search of lead-free perovskites, the replacement of lead with eco-friendly elements like bismuth ($Bi^{3+}$) and sodium ($Na^+$) may be a good strategy. Bismuth has very similar electronic properties as that of lead so it gives very efficient perovskites. Therefore, double perovskites containg Bi element $X_2NaBiCl_6$ (X=Cs, Rb, K) are explored here in terms of the structural, opto-electronic, and thermoelectric properties using the first-principles approach. The materials are attractive for optoelectronic applications such as ultraviolet sensors and detectors due to the prominent absorption peaks in ultraviolet region (10eV-30eV), low reflection (5-7%), and high optical conductivity ($\sim 10^{15}$ $sec^{-1}$). In addition to the alluring optoelectronic features, the compounds under study have figure of merit being close to unity so these are also promising for thermoelectric applications

**Keywords:** Double Perovskites; Density Functional Theory (DFT); Quantum ESPRESSO; BoltzTraP, Optoelectronic Properties; Thermoelectric Properties;

1. **Introduction:**

The traditional power generation methods based on fossil fuel are becoming less favorite among the scientific community because of the shortage of fuel resources around the globe and the immense environmental pollution caused by these methods. In this scenario, non-traditional power generation methods like solar cells and thermoelectric generators have gained tremendous attention as an alternative to fossil fuel-based schemesto fulfill the energy needs for future generations without polluting the environment. Due to attractive optoelectronic and thermoelectric device applications, perovskites have been widely explored. This is due to some novel characteristics for example tunable band gaps, large absorption coefficients, small effective masses, high carrier mobility and charge diffusion, and high figure of merits [1,2]. Secondly, perovskites are widely available, exceptionally efficient, and inexpensive when compared to silicon, which is why the researchers are persuing perovskite for solar cells and other important relevant optoelectronic and thermoelectric applications. Perovskites have their usage not only in solar cells and thermoelectric generators, but also many others for instance light emitting diodes (LEDs), ultraviolet detectors, ultraviolet sensors, lasers, and energy conversion devices [3-7].

Although perovskites possess remarkable characteristics as mentioned earlier but most of the efficient perovskites explored yet contains the lead element which is very toxic and badly affects the reproductive and nervous systems of the human body [8]. Therefore, many experimental, as well as theoretical studies, were conducted to tackle this issue by replacing the lead element with some suitable element in such a way that the efficiency and stability may not be compromised.Lead was replaced by tin either partially or completely to minimize the toxic effects without compromising the efficiency [9-12], however, it faced stability issues due to the oxidation of $Sn^{2+}$ [13].Keeping in view the stability issues of tin-based perovskites [14] and the limited availability of other suitable divalent cations, the researchers supposed the replacement of divalent lead with a combination of trivalent and monovalent cations to form double perovskites with general formulae $X_2ZZ'Y_6$ where $X=Cs^+$, $Rb^+$, $K^+$, $Z=Na^+, Cu^+, Ag^+, Z'=Bi^{3+}, Sb^{3+}$ whereas $Y=Cl^-$, $Br^-$, $I^-$ etc.[15]. In this way, the lead-free double perovskite became an attractive option for various optoelectronic and thermoelectric applications. Particularly, bismuth-based double perovskites were found very stable and efficient for instance, $Cs_2AgBiBr_6$ which has been synthesized [16] is observed as a very promising material for photovoltaic [17], photocatalytic [18], and photodetector [19, 20] applications.

Inspired by such appealing properties of bismuth-based double perovskites, we have investigated here $X_2NaBiCl_6$ (X=Cs, Rb, K) using a first-principles approach in terms of the structural, electronic, optical, and thermoelectric properties. We have compared the important findings of $Cs_2NaBiCl_6$ with existing literature for example lattice constants and band gap which are in good agreement with the reported data [21] however, the thermoelectric properties of $Cs_2NaBiCl_6$ have not been published earlier. We have substituted Rb and K in place of Cs and hence the compounds $Rb_2NaBiCl_6$ and $K_2NaBiCl_6$ are going to be explored for the first time.

2. **Computational details:**

The electronic structure and optical characteristics of the double perovskites $X_2NaBiCl_6$ (X=Cs, Rb, K) were studied from first-principles approach. Quantum ESPRESSO [22], a plane wave pseudopotential density functional theory (DFT) based code, was used to accomplish this task. In this study, the exchange-correlation functional used for computations is a Perdew, Burke, and Ernzerhof (PBE) [23] which is based on generalized gradient approximation (GGA). First of all, the structures were optimized i.e. both volume and geometry optimizations were performed so that the optimum lattice constants and atomic positions may be obtained. Norm-conserving pseudopotentials with k-mesh 15×15×15 were used for self-consistent field (SCF) and band structure calculations. A more denser K-mesh of 25×25×25 was considered for the calculations of the density of state (DOS) profiles, projected density of states (PDOS) and optical parameters. Kinetic energy and charge density cutoffs for wave function were optimized having values of 50 Ry and 500 Ry, repectively. Degauss value of Gaussian smearing was 0.02 Ry and the electron convergence criteria was $1\times10^{-6}$ Ry. The thermoelectric properties were calculated from BoltzTraP [24] . This code relies on well known Boltzmann transport theory-based and it uses the results of SCF calculations to compute various thermoelectric parameters.

## 3. Results and discussion:
### 3.1 Structural Properties:

The primitive cell of the double perovskite $X_2NaBiCl_6$ (X=Cs, Rb, K) as displayed in Fig. 1 depicts that the studied compunds are face-centered cubic (FCC) in structure having space group $Fm\bar{3}m$ (225).

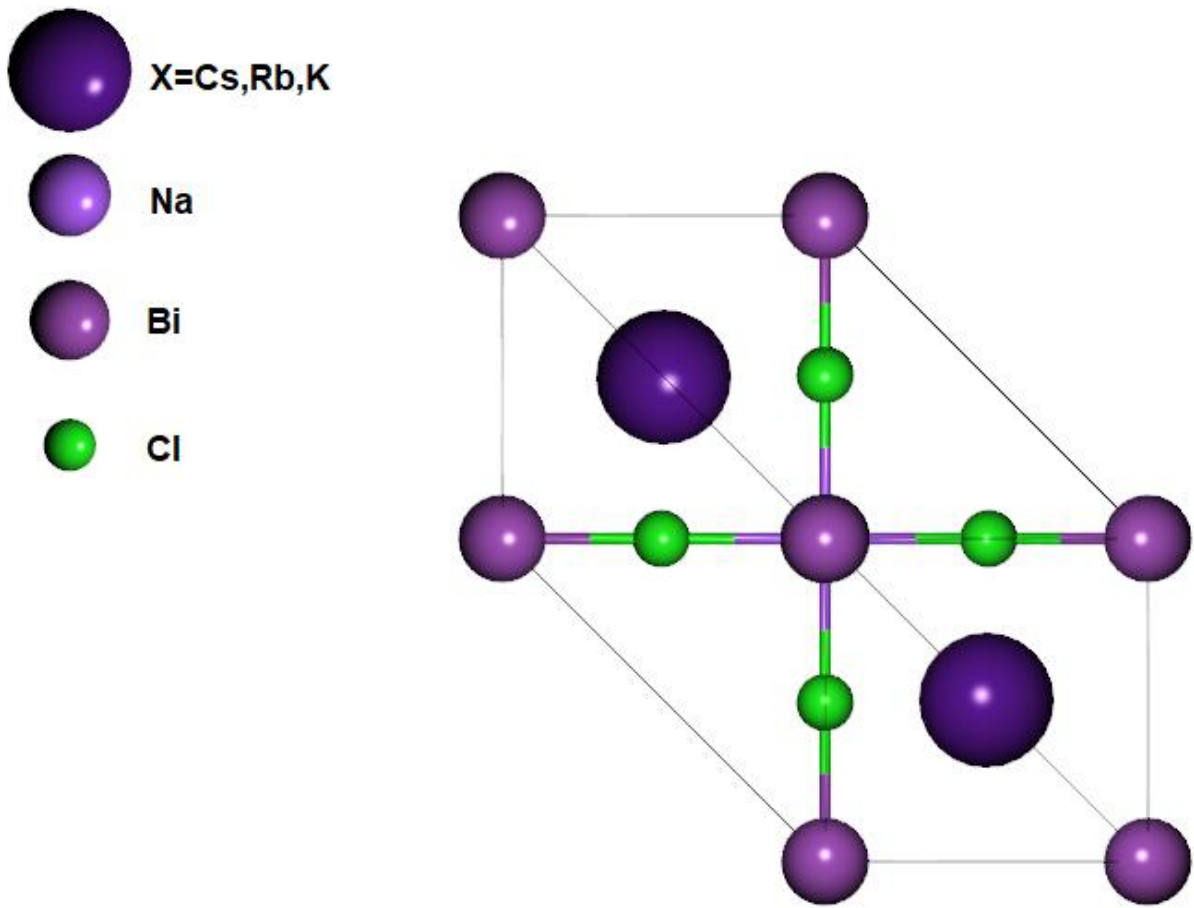

**Fig.1:** Pictorial representation of the primitive unit cell of $X_2NaBiCl_6$ (X=Cs, Rb, K).

The Wyckoff positions possessed by elements X (=Cs,Rb, and K), Na, Bi, and Cl are 8c, 4b, 4a, and 32f respectively. The standard Broyden, Fletcher, Goldfarb, and Shanno (BFGS) scheme is used for relaxation of the compunds. The energies are calculated against various volumes so that a minimum may be searched and hence, the optimized lattice constants are obtained by fitting the energy versus volume using Murnaghan's equation of state [25]. The details of the lattice constants obtained by above mentioned process of relaxation are tabulated in Table 1 and volume versus energy curves are presented in Fig. 2. By comparson with the published data, we came to know that the value of the lattice constant for $Cs_2NaBiCl_6$ agrees with literature [21] whereas $(Rb/K)_2NaBiCl_6$ has not been reported earlier.

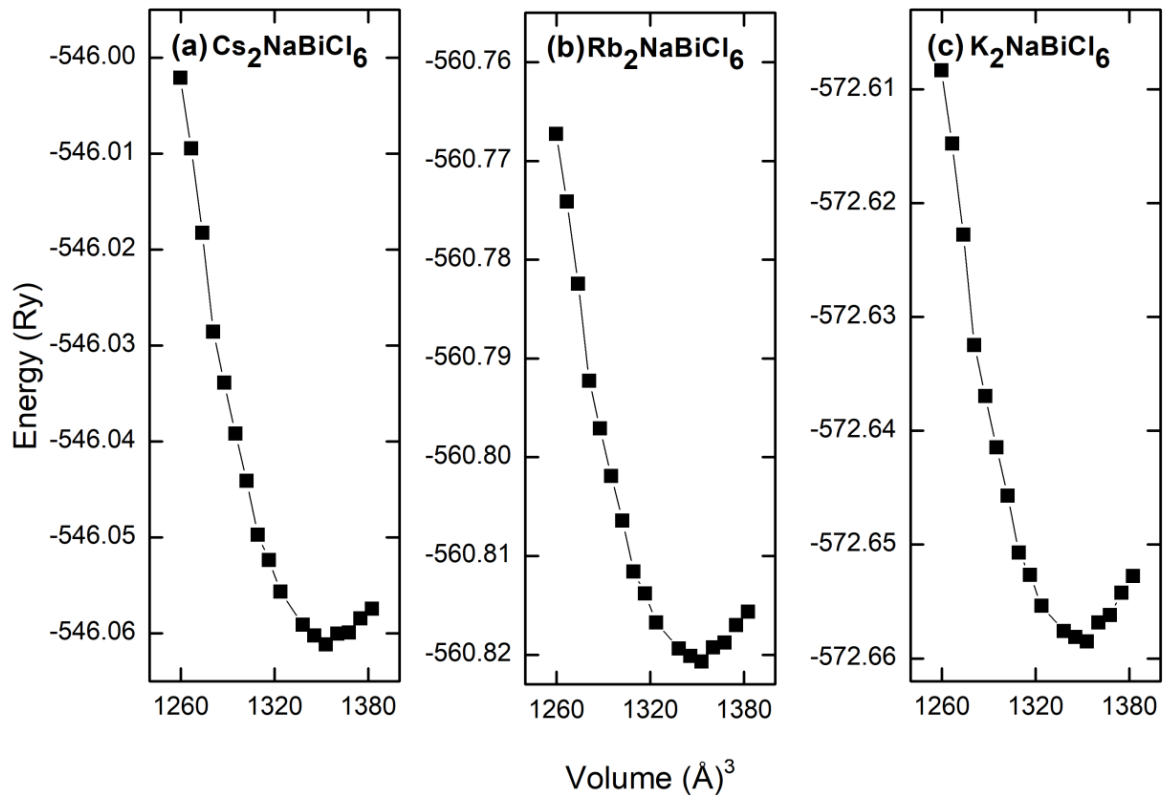

**Fig. 2:** Energy versus volume curves for $X_2NaBiCl_6$ (X=Cs, Rb, K).

Moreover, the assessment of the structure stability is also very cruitial point before discussing the further properties from the view point of applications in various domains. In case of perovskites, one of the very famous tools to asses the structural stability is Goldschmidt's tolerance ($\tau_o$) and octahedral factor ($\mu_o$) [26]. The factors $\tau_o$ and $\mu_o$ are given by equation (1-2).

$$\tau_o = \frac{r_X + r_{Cl}}{\sqrt{2}(r_B + r_{Cl})} \quad (1)$$

$$\mu_o = \frac{r_B}{r_{Cl}} \quad (2)$$

Here, $r_X$ is the ionic radius of Cs/Rb/K ions, $r_{Cl}$ is the radius of Cl ion whereas $r_B$ is the average of Na and Bi ionic radii.

The estimated values of $\tau_o$ were 0.87, 0.83, and 0.81 for $Cs_2NaBiCl_6$, $Rb_2NaBiCl_6$, and $K_2NaBiCl_6$ respectively and $\mu_o$ was constant with a value 0.69. The range of $\tau_o$ and $\mu_o$ values that is admitted stable for double perovskites are 0.813 <$\tau_o$< 1.107 and 0.377 <$\mu_o$< 0.895 [27], therefore, it is clear that the studied compounds have stable structures.

**Table 1: The estimated lattice constants for $X_2NaBiCl_6$(X=Cs, Rb, K).**

| Materials | Approximation | Lattice constants (Å) | References (Å) |
|---|---|---|---|
| $Cs_2NaBiCl_6$ | PBE (GGA) | 11.00 | 10.837(theoretical) [21] <br> 10.839(experimental) [28] |
| $Rb_2NaBiCl_6$ | PBE (GGA) | 11.04 | ----------- |
| $K_2NaBiCl_6$ | PBE (GGA) | 11.05 | ----------- |

### 3.2 Electronic Properties

Electronic properties of solids are very important and fundamental because of their influence on optical and thermoelectronic properties and in this scenario, the electronic band structure plays a very crucial role by dislosing disclose various applications of solids in devices like solar cells, thermoelectric devices, and transistors, etc. [29]. In particular, one can categorize a solid into metal, semimetal, and insulator by calculating band gaps. The electronic band structures of $X_2NaBiCl_6$ (X=Cs, Rb, K) are presented in Fig. 3(a-c) along the high symmetry path L-X-W-L-Γ-W. The band structure profiles reveal the indirect band gap nature of the semiconductors under question having valence band maxima (VBM) at high symmetry point W and conduction band minima (CBM) at high symmetry point L. The band gap values decrease if we replace Cs with Rb and further decrease while replacing with K. Table 2 contains the detailed information of the band gap values as well as the comparison with the available experimental/theoretical data. We can further clarify the electronic structures by analyzing the elemental and orbital contribution from the constituent elements in terms of DOS and PDOS. The DOS and PDOS profiles for $Cs_2NaBiCl_6$, $Rb_2NaBiCl_6$, and $K_2NaBiCl_6$ are displayed in Figs. 4-6. The group on the left of the dotted line representing the valence band (VB) whereas the other on the right of the dotted line is representing the conduction band (CB). In the case of $Cs_2NaBiCl_6$, the Cs-s, Cs-p, Na-p, and Cl-p have prominent peaks in VB however Bi has intermediate peaks in both VB and CB as shown in Fig.4 (a-e). Similarly, we can see that there are Rb-s and Na-s peaks in VB while Cl-p peaks are present in both VB and CB for $Rb_2NaBiCl_6$ as shown in Fig.5 (a-e). There are K-s, Cl-p, and Na-s peaks in VB whereas Cl-p and Bi-p of low intensity are there in CB as shown in Fig.6 (a-e) for the case of $K_2NaBiCl_6$.

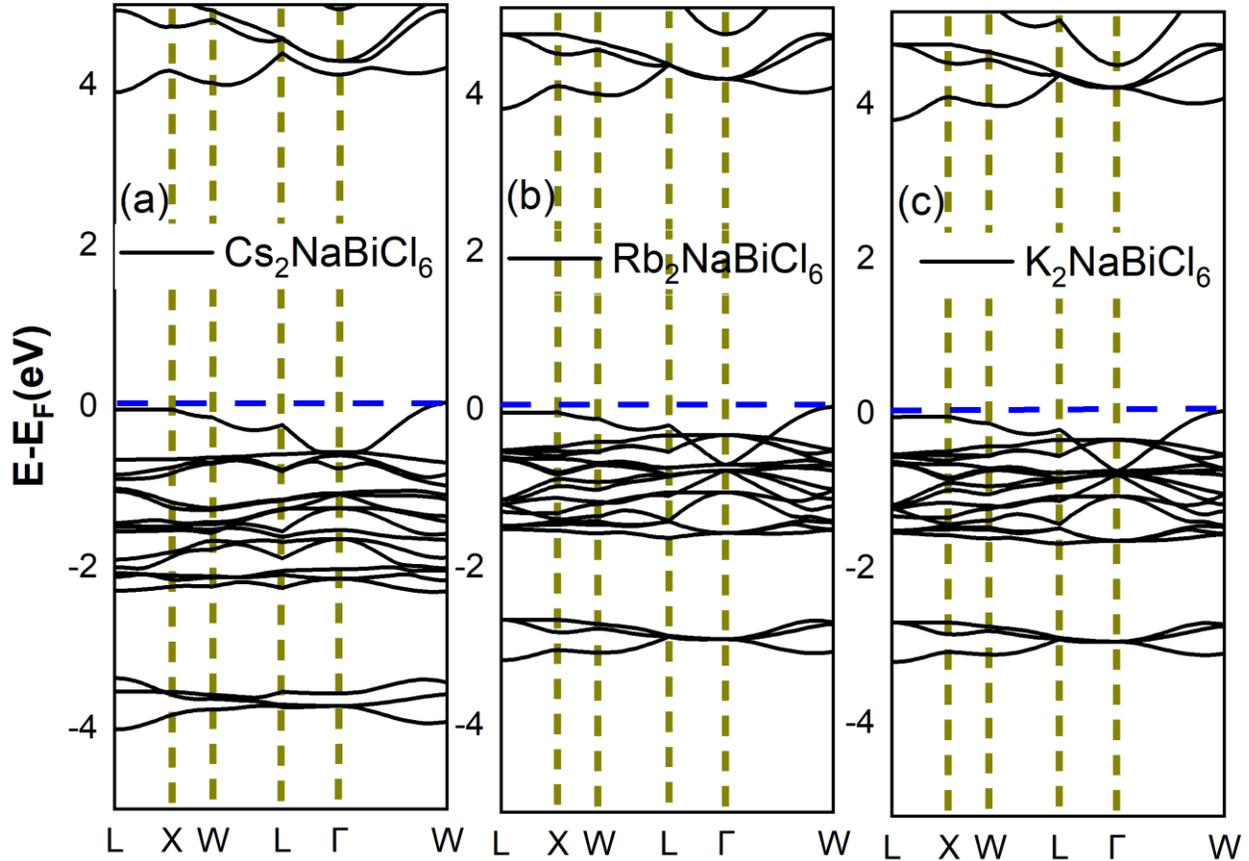

**Fig. 3:** Band structures of (a) $Cs_2NaBiCl_6$, (b) $Rb_2NaBiCl_6$, and (c) $K_2NaBiCl_6$.

**Table 2:** Calculated band gaps of $X_2NaBiCl_6$ (X=Cs, Rb, K).

| Compound | DFT Functional | Band gaps (eV) | References (eV) |
|---|---|---|---|
| $Cs_2NaBiCl_6$ | PBE | 3.91 | 3.73 [21] |
| $Rb_2NaBiCl_6$ | PBE | 3.79 | ----------- |
| $K_2NaBiCl_6$ | PBE | 3.75 | ----------- |

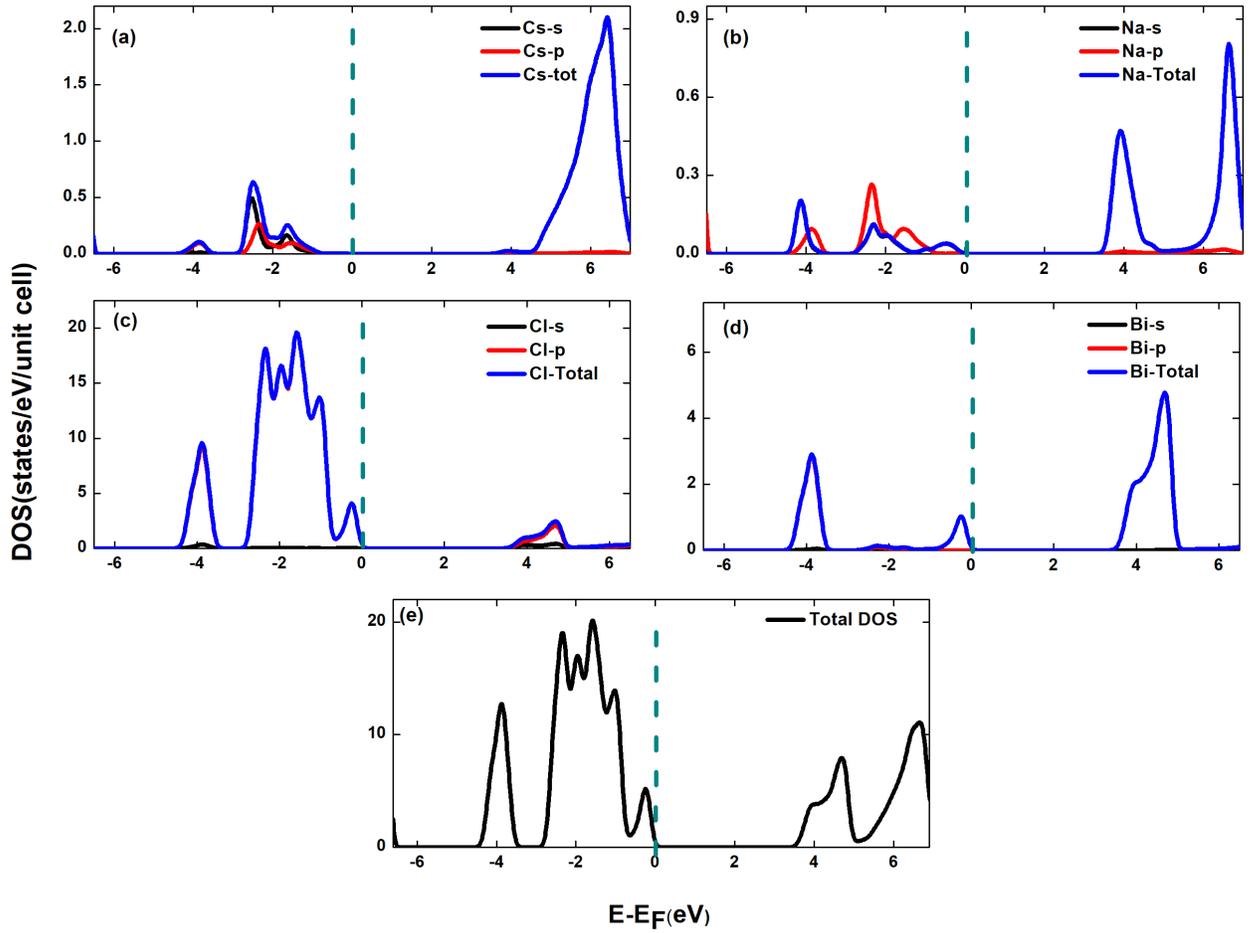

**Fig. 4:** Profiles of DOS and PDOS for $Cs_2NaBiCl_6$.

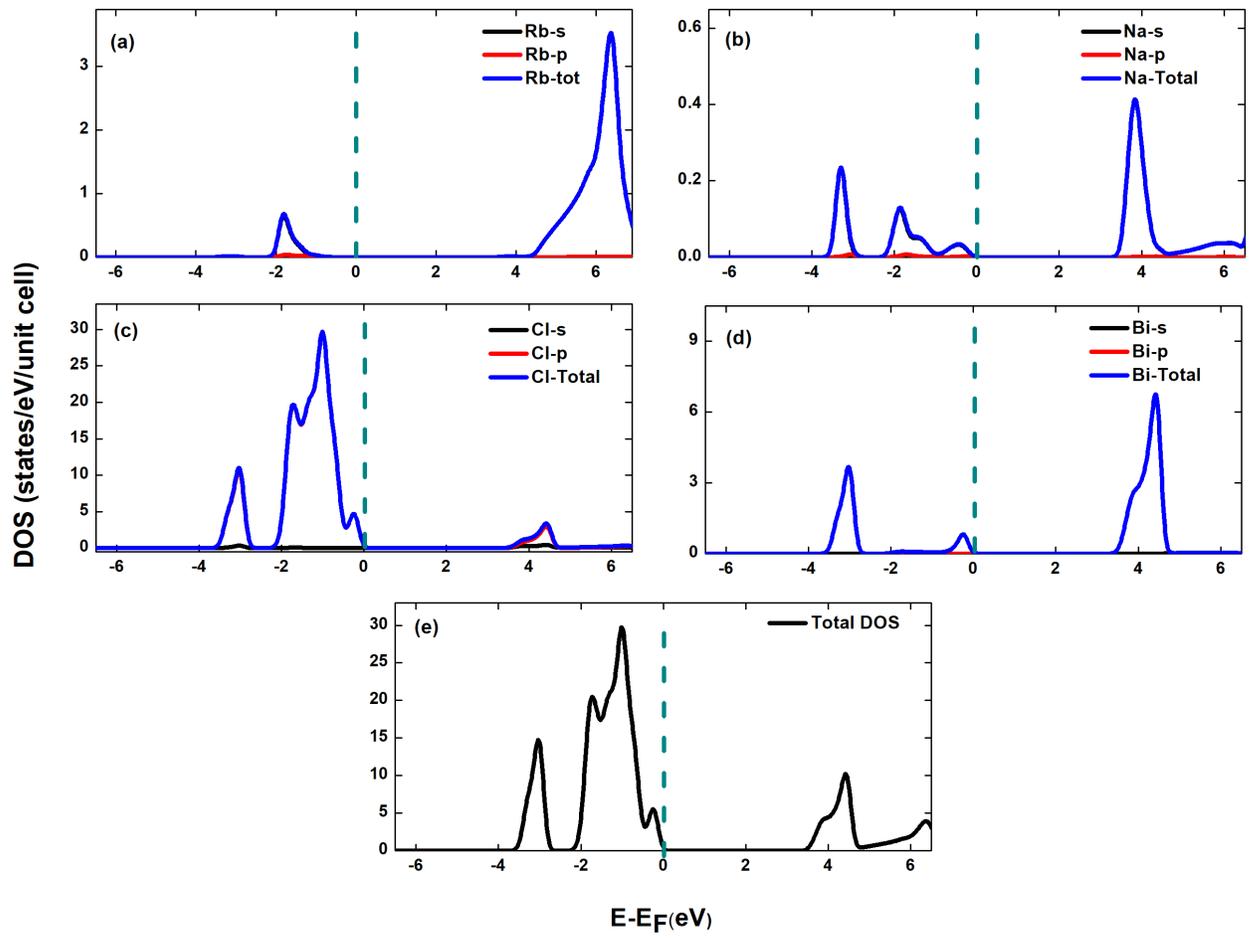

**Fig. 5:** Profiles of DOS and PDOS for Rb$_2$NaBiCl$_6$.

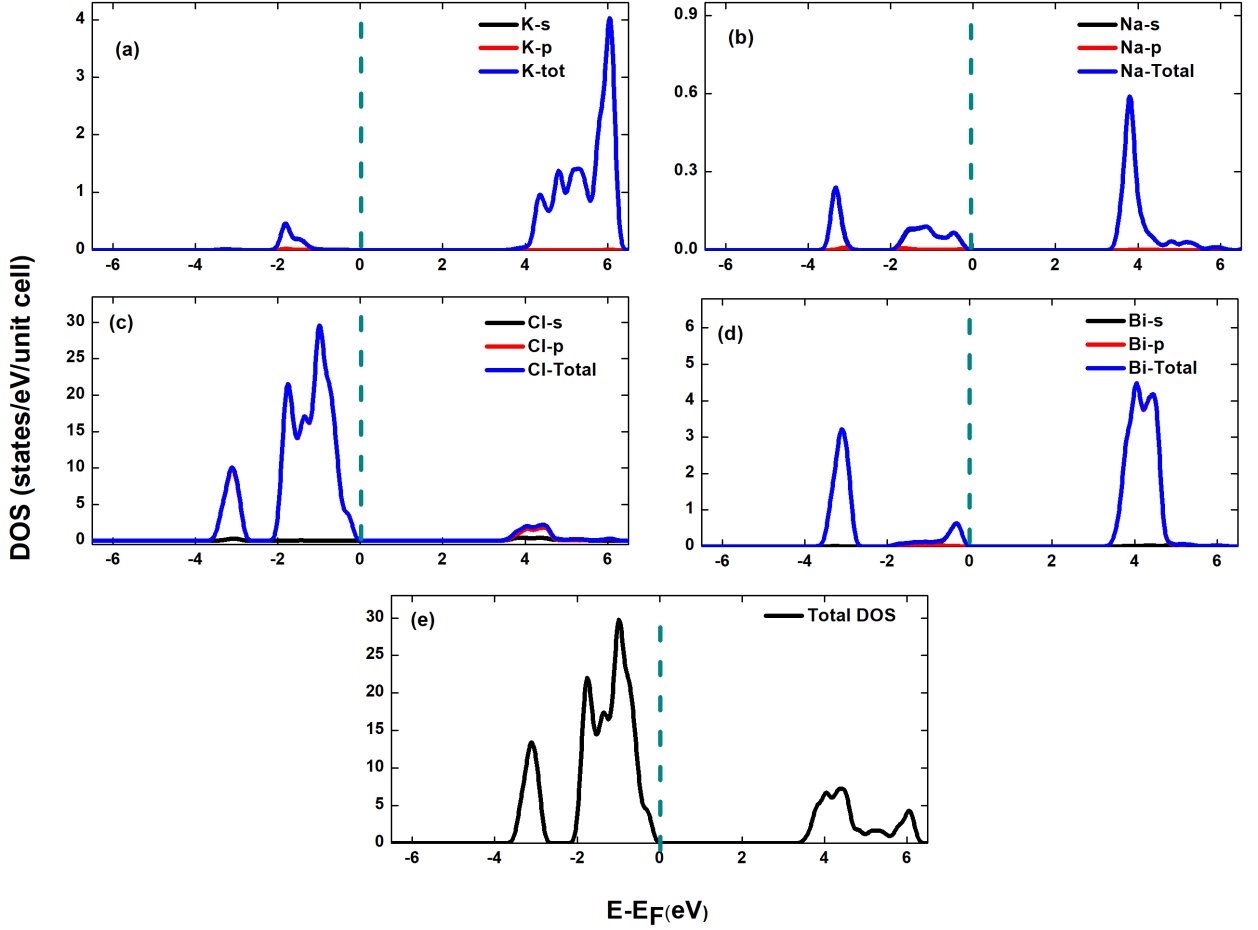

**Fig. 6:** Profiles of DOS and PDOS for $K_2NaBiCl_6$.

### 3.3 Optical Properties

Currently, we calculated complex dielectric function $\epsilon(\omega)$ which can be obtained from Ehrenreich and Cohen's relation [30]. Mathematically, it is a combination of real $\epsilon_{real}(\omega)$ and imaginary part $\epsilon_{imag}(\omega)$ and we can write it as follows:

$$\epsilon(\omega) = \epsilon_{real}(\omega) + i\epsilon_{imag}(\omega) \qquad (3)$$

The real part tells us about the stored energy whereas the imaginary part portrays the absorption behavior of the relevant material [31]. Not only this, we can also find other intresting optical parameters like optical absorption coefficient, percentage reflectivity, refractive index, extinction coefficient, optical conductivity, and energy loss function in order to examine the behavior of materials when exposed to light. Therefore, we calculated above mentioned parameters from the real and imaginary part of the dielectric function using equations (4-9).

In Fig.7 (a), we plotted $\epsilon_{real}(\omega)$ versus incident photon energy for a range of 0-20eV. It is evident that the static dielectric constant $\epsilon_{real}(0)$ are almost identical having a value of 2.57. The reasonable $\epsilon_{real}(0)$ peaks resides from 3.5eV to 7.5eV. Likewise, the Fig.7 (b) portrays that the significant peaks of $\epsilon_{imag}(\omega)$ reside between 4.5eV to 8eV.

Absorption coefficient $\alpha(\omega)$ given by equation (4), refers to the electronic transition from the VBM to CBM because of the energetically suitable photon absorption.

$$\alpha(\omega) = \frac{\sqrt{2}\omega\left[\{\epsilon_1^2(\omega)+\epsilon_2^2(\omega)\}^{1/2}-\epsilon_1(\omega)\right]^{1/2}}{c} \quad (4)$$

The visible peaks start from 4eV and the other prominent peaks reside in 5.4eV, 8.10eV, and 13.38eV.

Apart from absorption, each material has a certain level of reflection of incident photons. The measure of the amount of reflection in percentage is known as percentage reflectivity R% which can be calculated from equation (5).

$$R(\omega) = \frac{(n-1)^2+K^2}{(n+1)^2+K^2} \quad (5)$$

The R% in our case is very low i.e. 5-7% in the visible energy range (1.7eV-3eV), however, it reaches its maximum value of 55% at 9.4eV.

A dimensionless parameters refractive index $n(\omega)$ is measure of light bending capability so it is very useful parameter to categorize a material for various optoelectronic applications like waveguides, detectors, and solar cells [32]. It can be calculated using equation (6).

$$n(\omega) = \frac{\left[\{\epsilon_1^2(\omega)+\epsilon_2^2(\omega)\}^{1/2}+\epsilon_1(\omega)\right]^{1/2}}{\sqrt{2}} \quad (6)$$

Here, the value of the static refractive index is 1.58 for all three compounds under study and its maximum value of 2.5 occurs around 4.4eV. There is a direct relation between $n(\omega)$ and $\epsilon_{real}(\omega)$ i.e. $n(\omega) = \sqrt{\epsilon_1(\omega)}$[33].The the imaginary part of the refractive index given by equation(7) is commonly known as extinction coefficient $K(\omega)$.

$$K(\omega) = \frac{\left[\{\epsilon_1^2(\omega) + \epsilon_2^2(\omega)\}^{1/2} - \epsilon_1(\omega)\right]^{1/2}}{\sqrt{2}} \quad (7)$$

It can be seen from Fig. 7(f) that there is a similar trend of $K(\omega)$ and $\epsilon_{imag}(\omega)$. Another important parameter while exploring the optical properties of a material is optical conductivity $\sigma(\omega)$. It is directly related to the absorption coefficient and tells about the conduction of carriers upon the absorption of photons. It can be calculated by equation (8) [34].

$$\sigma(\omega) = \frac{\alpha(\omega) n(\omega) c}{4\pi} \quad (8)$$

The highest value of $4.79 \times 10^{15}$ sec$^{-1}$ was noted at 5.27eV. The other prominent peaks of optical conductivity $\sigma(\omega)$ in the range of $\sim 10^{15}$ sec$^{-1}$ are very attractive for various optoelectronic applications.

The loss function $L(\omega)$ given by equation (9), informs about the energy loss during the passage of photon from a material.

$$L(\omega) = \frac{\epsilon_2(\omega)}{\epsilon_1^2(\omega) + \epsilon_2^2(\omega)} \quad (9)$$

It has prominent peaks at 9.5eV and 14.05eV as it is clear from Fig.7 (h).

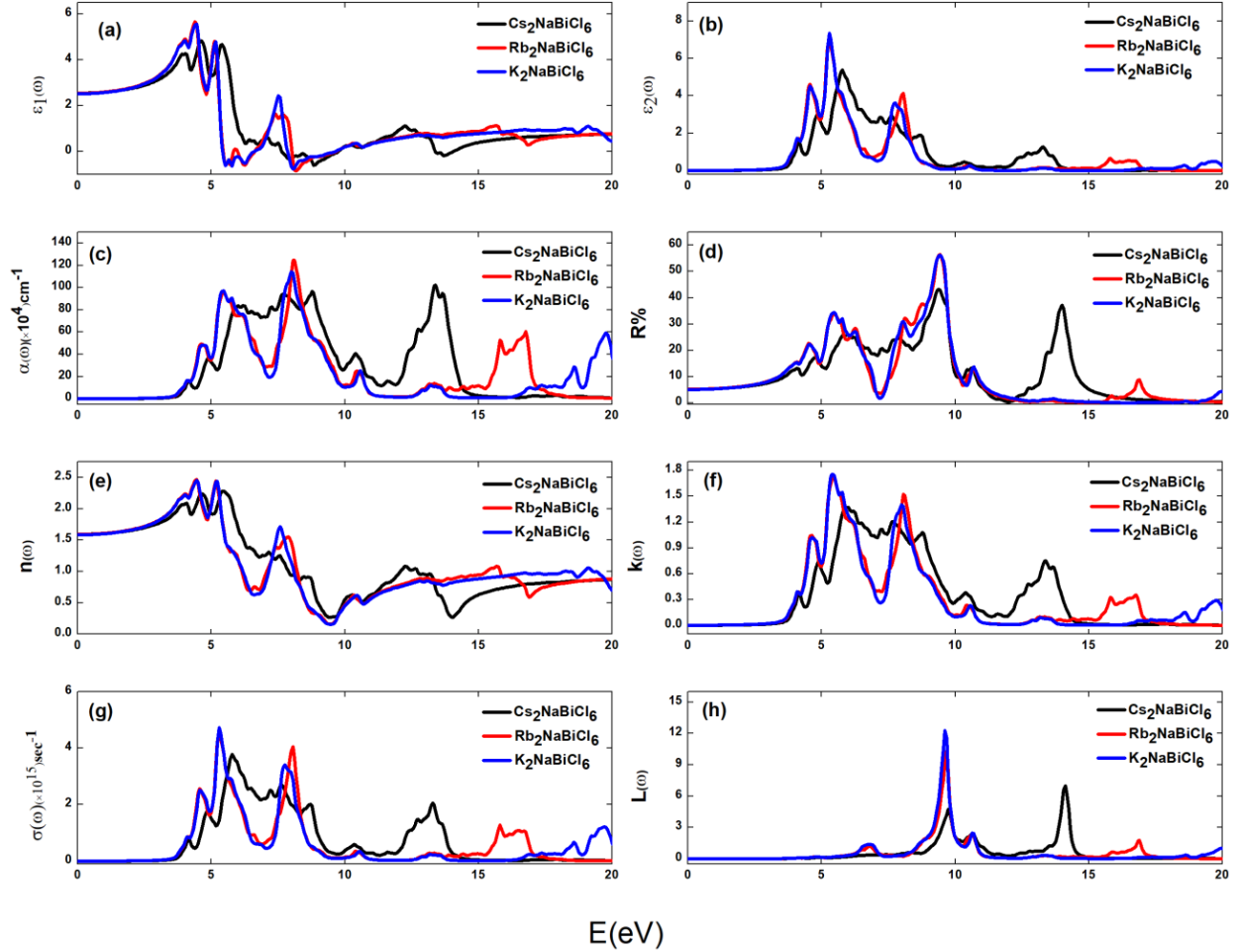

**Fig. 7:** Optical parameters (a) real part of dielectric function, (b) imaginary part of dielectric function, (c) absorption coefficient, (d) percentage reflectivity, (e) refractive index, (f) extinction coefficient, (g) optical conductivity, and (h) energy loss function for $X_2NaBiCl_6$ (X=Cs, Rb and K).

### 3.4 Thermoelectric Properties

The important thermoelectric parameters like electrical conductivity (σ), thermal conductivity (κ), thermopower (S), power factor (PF), and the electronic part of the figure of merit ($ZT_e$) for Bismuth-based double perovskites $X_2NaBiCl_6$ (X=Cs, Rb, K) were calculated.

Electrical conductivity refers to the mobilization of charges in response to the temperature gradient in the thermoelectric materials (TEMs). According to the classical [35], it can be calculated using equation (10).

$$\frac{\sigma}{\tau} = \frac{ne^2}{m^*} \qquad (10)$$

The symbols n, e and m* represents charge density, electric charge, and effective mass respectively. Temperature dependence of electrical conductivity can be interpreted mathematically using the following expression [36]:

$$\sigma(T,\mu) = \frac{1}{\Omega} \int \sigma_{\alpha\beta}(\epsilon) \left[ \frac{-\partial f_o(T,\mu,\epsilon)}{\partial \epsilon} \right] \partial \epsilon \qquad (11)$$

Where $\Omega$ is the unit cell volume, $\sigma_{\alpha\beta}(\epsilon)$ represents the transport energy distribution tensor, $\mu$ is the chemical potential whereas $\alpha, \beta$ are the indices of the energy distribution tensor. The variations of σ/τ with absolute temperature (T) near the Fermi level are graphically represented in Fig. 8 (a). For all compounds, the values of electrical conductivity per unit relaxation time (σ/τ) increase with temperature, reaching maximum $1.96 \times 10^{18}(\Omega ms)^{-1}$, $1.38 \times 10^{18}(\Omega ms)^{-1}$, and $0.34 \times 10^{18}(\Omega ms)^{-1}$ for $Cs_2NaBiCl_6$, $Rb_2NaBiCl_6$, and $K_2NaBiCl_6$ respectively.

Thermal conductivity (κ) refers to energy transportation in the form of heat in response to the temperature gradient. The value of κ contains two components. The first is due to electronic transport commonly known as the electronic part of thermal conductivity ($κ_e$) and second due to lattice vibrations (phonons) which is known as lattice component of thermal conductivity ($κ_l$).

$$κ = κ_e + κ_l \qquad (12)$$

The value of $κ_l$ has an inverse relation with absolute temperature [37, 38] thus it can be ignored at elevated temperatures. Therefore, we have calculated the thermal conductivity (electronic part) per unit of relaxation time ($κ_e/\tau$) for $X_2NaBiCl_6$ (X=Cs, Rb, K) that is displayed in Fig. 8 (b). The variation of $κ_e/\tau$ with respect to the temperature has a similar increasing trend as that of σ/τ. This phenomenon occurs due to the direct relation between electrical and thermal conductivity as described by the Wiedemann-Franz law [39]. The largest values of $κ_e/\tau$ were $1.96 \times 10^{14}$ W/mks, $1.57 \times 10^{14}$ W/mks, $0.39 \times 10^{14}$ W/mks for $Cs_2NaBiCl_6$, $Rb_2NaBiCl_6$, and $K_2NaBiCl_6$ respectively.

The relaxation time is kept constant ($\tau \sim 10^{-14} s$) under the constant relaxation time approximation, which is very successful in predicting the various thermoelectric properties [40].

Thermo-power (S) refers to the voltage induced in response to the temperature gradient. It can be calculated using equation (13) [36].

$$S(T, \mu) = \frac{1}{eT\Omega\sigma_{\alpha\beta}(T,\mu)} \int \sigma_{\alpha\beta}(\epsilon)(\epsilon - \mu) \left[\frac{-\partial f_o(T,\mu,\epsilon)}{\partial \epsilon}\right] \partial \epsilon \quad (13)$$

The studied material hasa sharpgradient in the Seebeck coefficient in the temperature 0-200K and after that, the values attain a uniform trend with temperature. The recorded maximum values for $Cs_2NaBiCl_6$, $Rb_2NaBiCl_6$, and $K_2NaBiCl_6$ are -1347µV/K, -2549µV/K, and -2740µV/K respectively. All compounds under study show negative values which is evidence of n-type semiconducting behavior.

Another important thermoelectric parameter that determines a thermoelectric device's effectiveness is the power factor (PF). Mathematically:

$$PF = S^2\sigma \quad (14)$$

The maximum values of PF were $2.16 \times 10^{11}$ W/mk$^2$s, $1.78 \times 10^{11}$ W/mk$^2$s, and $0.43 \times 10^{11}$ W/mk$^2$s for $Cs_2NaBiCl_6$, $Rb_2NaBiCl_6$, and $K_2NaBiCl_6$ respectively.

Figure of merit (ZT), a dimensionless quantity, is a measure of conversion efficiency of TEMs. It is defined by equation (15).

$$ZT = \frac{S^2\sigma T}{\kappa} \quad (15)$$

It is evident that reasonable ZT requires large Seebeck coefficients, and big ration between electrical to thermal conductivity. The TEMs having ZT equal to one or greater than one are accepted as appropriate materials for thermoelectric device applications. It is evident from Fig. 9 that the ZT values have a decreasing trend that can be interpreted by the temperature dependence of the band gap. The relation between band gap and temperature is given by equation (16)[41].

$$E_g(T) = E_g(0) - \frac{\alpha T^2}{T+\beta} \quad (16)$$

The highest values of ZT are observed as 1.00 in the case of $Cs_2NaBiCl_6$ and 0.99 for both $K_2NaBiCl_6$ and $Rb_2NaBiCl_6$. The large values of electrical conductivities and Seebeck coefficients and values of ZT close to unity make these materials very attractive for thermoelectric device applications.

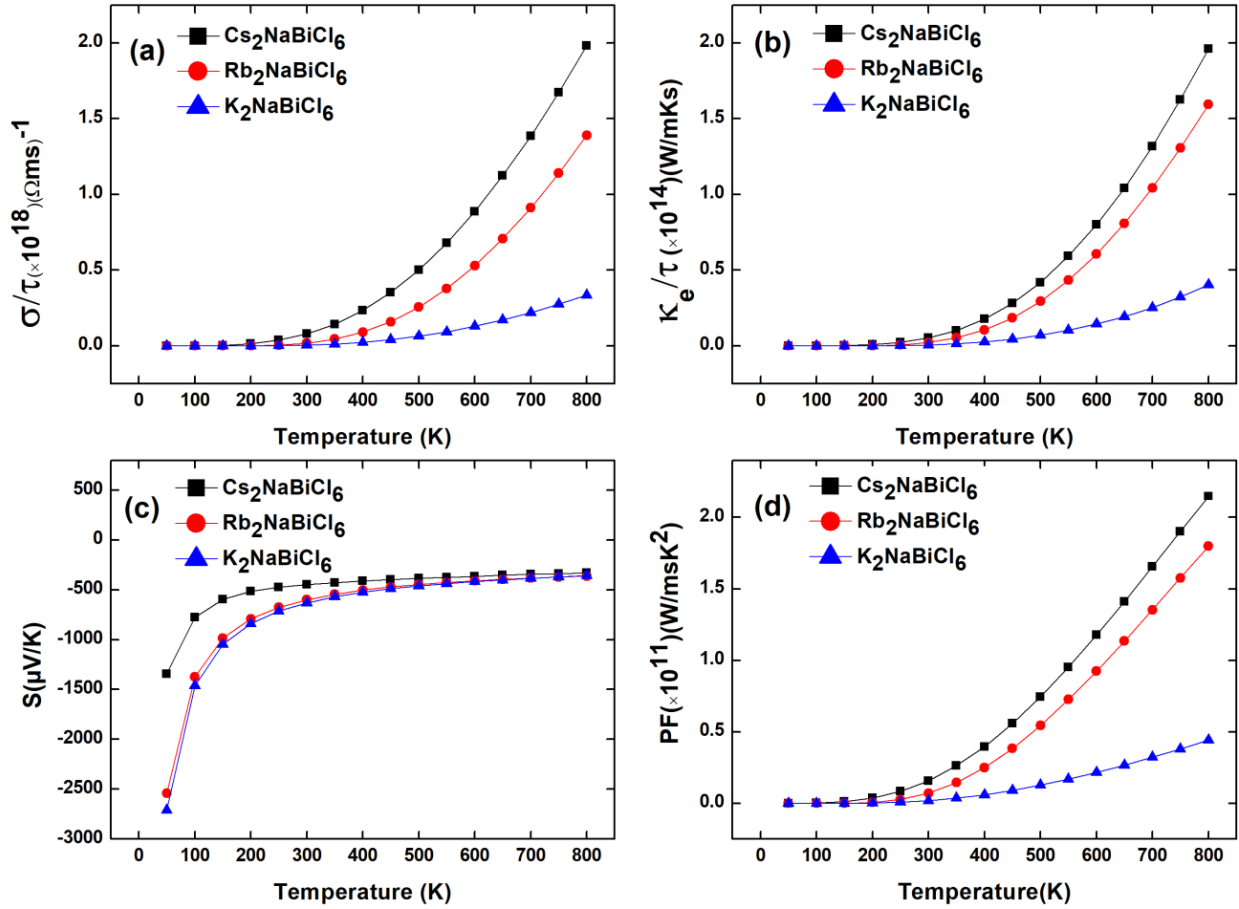

**Fig. 8:** Thermoelectric parameters (a) electrical conductivity, (b) thermal conductivity (electronic part), (c) Seebeck coefficient, and (d) Thermo-power for $X_2NaBiCl_6$ (X=Cs, Rb and K)

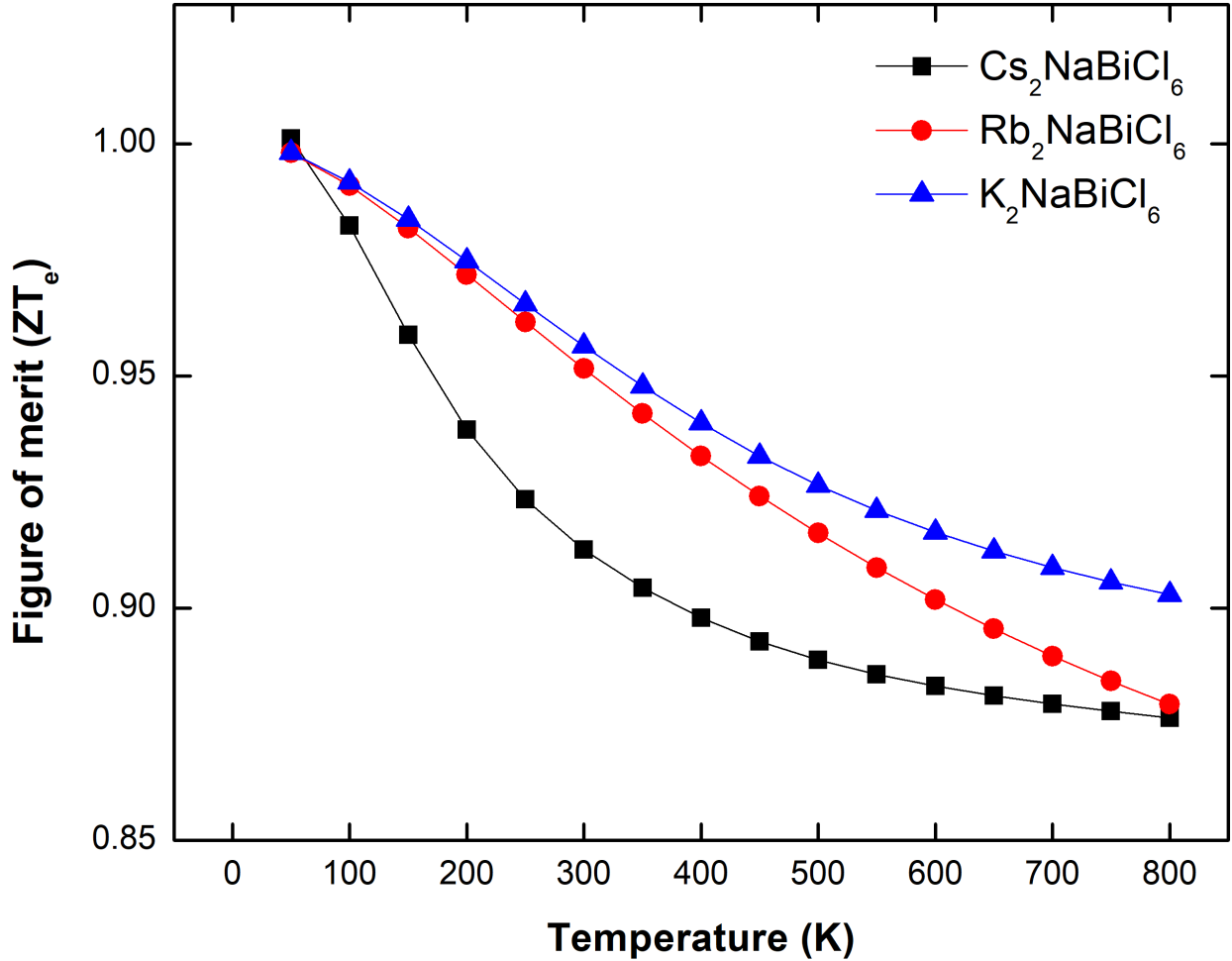

**Fig. 9:** Thermoelectric figure of merit w.r.t. temperature.

4. **Conclusion**

The green energy harvesting double perovskites based on bismuth $X_2NaBiCl_6$ (X=Cs, Rb, K) are investigated using DFT-based code Quantum ESPRESSO within generalized gradient approximation in terms of the structural, electronic, and optical properties. The thermoelectric parameters are evaluated using Boltzmann transport theory-based code BoltzTraP coupled with Quantum ESPRESSO. All three compounds under investigation have FCC structures and their structural stability is assessed from Goldschmidt's tolerance and octahedral factors. The compounds possess indirect band gaps of 3.91eV, 3.79eV, and 3.75eV for $Cs_2NaBiCl_6$, $Rb_2NaBiCl_6$, and $K_2NaBiCl_6$ respectively. The reasonable absorption peaks in visible and ultraviolet energy regions, low reflection coefficients (5-7%), the static refractive indices with a value of 1.58, and high optical conductivity (~$10^{15} sec^{-1}$) portray the novelty of the compounds for multiple optoelectronic applications like ultraviolet sensors, and detectors etc. Moreover, concerning the thermoelectric features, high electric conductivities in the range $\sim 10^{18}(\Omega ms)^{-1}$, high negative Seebeck coefficients, large power factors in the range of $\sim 10^{11} W/mk^2 s$, and figure of merits close to unity are observed. These attractive results provide the theoretical evidence for producing lead-free double perovskites with promising optoelectronic and thermoelectric device applications.

**Acknowledgment**

The Centre for High Performance Computing (CHPC-MATS1424), Cape Town, South Africa is acknowledged by the authors to facilitate in computational resources.